\documentclass
[aps,twocolumn,prl,showpacs,preprintnumbers,amsmath,amssymb]{revtex4}
\usepackage{natbib}
\usepackage[dvips]{graphicx}
\usepackage{dcolumn}
\usepackage{bm}
\usepackage{times}
\usepackage{mathptmx}
\usepackage[usenames]{color}

\begin{document}
\draft

%\title{Theory of photo-induced dynamics in double-exchange model}
\title{Dynamical coupling and separation of multiple degrees of freedom in a photoexcited double-exchange system}
\author{Y.~Kanamori$^1$, H.~Matsueda$^{2}$, and S. Ishihara$^1$} 
\address{$^1$Department of Physics, Tohoku University, Sendai 980-8578, Japan}
\address{$^2$Sendai National College of Technology, Sendai, 989-3128, Japan}
\date{\today}

\begin{abstract}
We present a theory of ultrafast photo-induced dynamics in 
a spin-charge coupled system, motivated by pump-probe experiments in perovskite manganites. 
A microscopic picture for multiple dynamics in spin and charge degrees is focused on.  Real-time simulations are carried out by two complimentary methods. 
%the exact-diagonalization and Hartree-Fock methods. 
Our calculation demonstrates that electron motion governs a short-time scale where charge and spin dynamics are combined strongly, while, in a long-time scale controlled by spin relaxation, 
charge sector does not follow remarkable change in spin sector. 
Present results are in contrast to a conventional double-exchange picture in equilibrium states. 
\end{abstract}


%71.30.+h 	Metal?insulator transitions and other electronic transitions
%78.47.J-    Ultrafast pump/probe spectroscopy (< 1 psec)
%78.20.Bh 	Theory, models, and numerical simulation
%71.10.-w 	Theories and models of many-electron systems

\maketitle
\narrowtext

Ultrafast photo-control of electronic and magnetic structures has attracted much attention for a long time from 
view points of fundamental physics and technological application. 
In correlated electron systems with multiple degrees of freedom, i.e. charge, spin, orbital and lattice,  
a stable equilibrium phase is determined by a subtle balance of several interactions between them~\cite{book}. 
In a barely stable state at vicinity of phase boundary, 
a gigantic change in electronic structure is triggered by an optical pump pulse. 
Recently developed several time-resolved experiments 
%, e.g. the optical absorption, the photoemission spectroscopy, the x-ray diffraction and so on, 
enable us to 
access directly to the photo-dynamics of the multiple degrees of freedom~\cite{JPSJ}. 
%, and uncover some aspects of photo-induced phenomena~\cite{JPSJ}. 

Exotic equilibrium phenomena in correlated electron systems, such as, high-Tc superconductivity, colossal magnetoresistance (CMR), multiferroics, and others, are often attributed to strong correlation between multiple degrees of freedom.  
It is not quite trivial whether this naive strong-coupling picture 
between multiple degrees is applicable to the photo-induced dynamics or not. 
One example is seen in perovskite manganites~\cite{miyano,fiebig,averitt,rini}.  
A key issue is a coupling between spin and charge. 
The CMR effects are resulted from a subtle balance of the charge ordered (CO) insulating phase associated with the antiferromagnetic (AFM) order and the ferromagnetic metallic (FM) phase, 
and are addressed in strong coupling between localized spins and conductive electrons~\cite{book}.  
By irradiation of a femto-second laser pulse in the CO insulating phase near the boundary, 
a transient metallic state expected from a reflectivity change appears within 100fs. 
While a macroscopic magnetization suggested from the magneto-optical Kerr rotation increases with the time of few ps \cite{ogasawara1,mcgill,ogasawara,miyasaka,matsubara}. 

These results imply that 
changes in spin and charge sectors, which are recognized to be strongly correlated in manganites,  
are not always accompanied with each other in photo-excited state. 
This dynamical coupling and separation of multiple degrees of freedom
is not only a clue for an overall understanding 
of the optical-control electronic phases in correlated system. 
This also provides a key to uncover exotic phenomena in equilibrium state; 
roles of the multiple degrees are able to be identified separately.
In spite of the progressive researches, however, 
a microscopic picture of multiple dynamics in photo-excited states has not been clarified yet.

In this Letter, motivated from the pump-probe experiments in manganites, 
we present a theory of the photo-induced dynamics in a spin-charge coupled system 
described by the double-exchange (DE) model. 
We focus on dynamical coupling and separation between spin and charge. 
To attack this issue, two complimentary methods, the exact-diagonalization (ED) and the Hartree-Fock (HF) methods, are adopted. 
Calculations demonstrate that electron motion governs the short-time scale where charge and spin sectors change cooperatively. On the other hand, in the long-time scale controlled by spin relaxation, charge sector does not follow change in the long-range spin correlation.
%The present results provide a microscopic interpretation for the observed 
%multiple time scales in $\Delta R$ and $\Delta \theta$. 

Let us introduce the extended DE model, 
\begin{eqnarray}
{\cal H}_{DE}&=&- \alpha t \sum_{\langle i j \rangle a} \left ( c_{i a}^\dagger c_{j a} +H.c. \right ) 
-J_H \sum_i {\bf s}_i \cdot {\bf S}_i
\nonumber \\
&+&U \sum_i n_{i \uparrow} n_{i \downarrow}
+V \sum_{\langle ij \rangle } n_i n_j
+J_S \sum_{\langle ij \rangle }{\bf S}_i \cdot {\bf S}_j , 
\label{eq:hamiltonian}
\end{eqnarray}
where $c_{i a}$ is the annihilation operator for the conduction electron at site $i$ with spin $a(=\uparrow, \downarrow)$, and ${\bf S}_i$ is the operator for the localized spin. 
We introduce the number operator $n_i(\equiv \sum_a n_{i a}=\sum_a c_{i a}^\dagger c_{i a})$ and the spin one ${\bf s}_i=\frac{1}{2} \sum_{a b} c_{i a}^\dagger {\bf \sigma}_{a b} c_{i b}$ with the Pauli matrices ${\bf \sigma}$ for the conduction electrons. 
The first and second terms describe the electron transfer, $\alpha t$, and the Hund coupling, $J_H$, respectively. 
The on-site Coulomb repulsion $U$, the nearest-neighbor (NN) one $V$, and the AFM superexchange interaction $J_S$ between the localized spins are also taken into account. 
%The $e_g$ orbital degree of freedom in manganites is not considered, for simplicity. 
All energy and time parameters are given as a unit of $t$ and $t^{-1}$, respectively. 
A magnitude of the transfer integral is changed by changing $\alpha$ from one. 
For a typical value of $t$ in manganite, 
a unit of time, $t^{-1}$ ,corresponds to 0.5fs$\sim$1fs.  

To analyze this model, two complimentary methods, the ED and HF ones, are utilized. 
In the ED method, the time-dependence of the wave function and the excitation spectra are able to be obtained exactly. On the other hand, in the HF one, simulations in a large system size and long time distance are possible, and the spin relaxations are taken into account.  

We utilize, in the ED method, the time-dependent and dynamical density-matrix renormalization group (DMRG) and Lanczos methods~\cite{matsueda}. One-dimensional clusters of system size $N_L=L(\leq 13)$ with the open-boundary condition are adopted. 
Electron number is $N_{ele}=(L+1)/2$ corresponding to the 1/4 filling. For simplicity, amplitude of the localized spin is set to be $1/2$. We take a damped oscillator form for the vector potential of the pump photon as $A_{pump}=A_0 e^{-i \omega_0 \tau- \gamma_0|\tau|}$ at time $\tau$ with frequency  
$\omega_0$ and a damping constant $\gamma_0$. 
A center of the wave packet is defined by $\tau=0$. 
The wave function for the one-photon absorbed state at time $\tau (> \gamma_0^{-1})$ is derived by the first order perturbation with respect to $A_{pump}$~\cite{photon}:    
\begin{eqnarray}
| \Psi(\tau) \rangle = {\cal N} e^{- i {\cal H}_{DE} \tau}
 \left [ \frac{\gamma_0}{ \left ( \omega_0-{\cal H}_{DE}+E_{0}\right )^2 + \gamma_0^2} \right ] 
 j | 0 \rangle , 
\label{eq:psitau}
\end{eqnarray}
where $j=i \alpha t\sum_{\langle ij \rangle a } (c_{ia}^\dagger c_{j a}-H.c.)$ is the current operator, $|0 \rangle $ is the wave function before pumping, $E_0=\langle 0 |{\cal H}_{DE}|0 \rangle$, and ${\cal N}$ is a normalization factor. 
%The static and dynamical quantities at time $\tau$ are calculated by using this wave function. 
Transient excitation spectra are calculated by the linear-response theory. For example, the one-particle excitation spectra are given by a sum of the electron and hole parts, $A(q, \omega)=A^{ele}(q, \omega)+A^{hole}(q, \omega)$, where the first term is obtained as \cite{approxi1} 
\begin{equation}
A^{ele}(q, \omega)= -\frac{1}{\pi} \sum_a 
{\rm Im} \langle c_{-q a}^\dagger \frac{1}{-\omega-{\cal H}_{DE}+E+i \gamma} c_{q a}  \rangle . 
\label{eq:aqw}
\end{equation}
We introduce the Fourier transform (FT) of the operator, $c_{q a}$, the energy $E=\langle {\cal H}_{DE} \rangle$, and a damping factor $\gamma$.  A bracket $\langle \cdots \rangle$ implies the expectation value with respect to $ | 0 \rangle $ in the case before pumping, and that with $| \Psi(\tau) \rangle$ after pumping. 
The pseudo-momentum in the open boundary condition is defined as 
$O_{q}=\sqrt{2/(L+2)} \sum _i \sin (q r_i ) O_i$
where $O_i$ ($O_{q}$) is a physical quantity in the real (momentum) space.  

In the HF method, the mean-field (MF) decoupling is introduced in the many-body terms in ${\cal H}_{DE}$, such as $n_{i \uparrow} n_{i \downarrow} \rightarrow \langle n_{i \uparrow} \rangle n_{i \downarrow}
+ n_{i \uparrow} \langle n_{i \downarrow} \rangle 
-\langle c_{i \uparrow}^\dagger c_{i \downarrow} \rangle  c_{i \downarrow}^\dagger c_{i \uparrow}
- c_{i \uparrow}^\dagger c_{i \downarrow} \langle  c_{i \downarrow}^\dagger c_{i \uparrow}  \rangle
+ const.$ where $\langle n_{i \uparrow} \rangle $ is a site-dependent MF. The two-dimensional $N_L=L \times L$ 
$(L=6-10)$ site cluster with the periodic-boundary condition is adopted. 
The electron number is $N_{ele}=L^2/2$. 
%and ${\bf S}_i \cdot {\bf S}_j \rightarrow \langle {\bf S}_i \rangle \cdot {\bf S}_j +{\bf S}_i \cdot \langle {\bf S}_j \rangle+const. 
The localized spins are treated as classical vectors with amplitude of $3/2$. 
The pump-photon irradiation is simulated as an electronic excitation from the highest occupied HF level to the lowest unoccupied one without changing spin. 
The time evolution of the electronic wave function is obtained by the time-dependent HF scheme~\cite{satoh,miyashita}; 
\begin{eqnarray}
\phi^{(l)}(\tau+\Delta \tau) =T \exp \left [    -i \int_\tau^{\tau+\Delta \tau}  {\cal H}_{DE}^{HF} (\tau') d \tau' \right ]
\phi^{(l)}(\tau) , 
\end{eqnarray}
where $\phi(\tau)^{(l)}$ is the $l$-the HF wave function, and  ${\cal H}_{DE}^{HF}(\tau')$ is the HF Hamiltonian for Eq.~(\ref{eq:hamiltonian}). Dynamics of the localized spins is described by the Bloch-type equation: 
\begin{equation}
\partial_\tau {\bf S}_i=-{\bf H}_i \times {\bf S}_i+\partial_\tau {\bf S}_i|_{relax} , 
\label{eq:bloch}
\end{equation}
where ${\bf H}_i$ is the MF at site $i$ 
defined by 
${\bf H}_i=J_H\langle {\bf s}_i\rangle+J_S\sum_\delta  {\bf S}_{i+\delta}$. 
The last term in Eq.~(\ref{eq:bloch}) 
%$\partial_\tau {\bf S}_i(\tau)|_{relax}$ 
represents the spin relaxation and dephasing which will be explained later. 

\begin{figure}[tb]
\begin{center}
\includegraphics[width=1.05\columnwidth,clip]{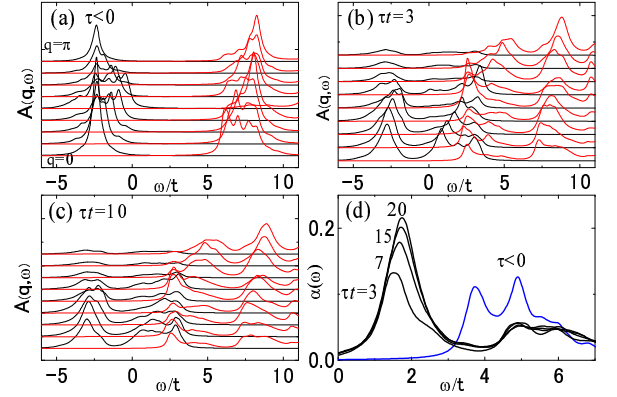}
\caption{(color online) 
(a) One-particle excitation spectra before pumping, (b) those at $\tau t=3$, and (c) those at $\tau t=10$.  
Momenta are changed from $\pi$ to 0 for the data from the top to the bottom.  
%Bold and thin lines are for the electron and hole parts, respectively. 
(d) Time-dependence of optical absorption spectra. 
%Bold curve is for the spectra before pumping. 
%Parameter values are chosen to be $U=10$, $J_H=8$, $V=5$, $J_S=0.4$, $\omega_0=3.72$ and $\gamma_0=0.4$ as a unit of $t$. 
%System size is $L=9$ with $N_{ele}=5$. 
}
\label{fig:fig1}
\end{center}
\vspace{-0.6cm}
\end{figure} 
First, we show the results obtained by the ED method. 
The parameter values are chosen to be $U=10$, 
$J_H=8$, $V=5$, $J_S=0.4$, $\omega_0=3.72$ and $\gamma_0=0.4$. 
We take $\alpha=1$ except for the results in 
the inset of Fig.~\ref{fig:fig3}. 
Somewhat larger energy parameter values than the realistic ones for manganites are
used to reproduce the AFM/CO insulating ground-state in one-dimension. 
Thus, our analyses in the ED method are restricted to qualitative properties in the photo-induced dynamics. 
%The ground state before pumping is the CO insulator with the ferrimagnetic spin 
%structure where the localized spins are aligned antiferromagnetically and the 
%total spin  quantum number is $S^{tot}=\frac{1}{2}+\frac{L+1}{4}$. 
First we focus on the charge dynamics. 
An insulating gap and a tendency of the Brillouin-zone doubling shown in $A(q,\omega)$  
[see Fig.~\ref{fig:fig1}(a)] 
imply an alternating charge alignment. 
%and the static spin- and charge-correlation functions (not shown in the figure). 
Just after photon pumping [see Fig.~\ref{fig:fig1}(b)], a photo-carrier band appears inside of the CO gap in $A(q, \omega)$. 
%The spectral intensities in the upper (lower) band around $q=0$ $(q=\pi)$ are reduced. 
%Although the upper, lower and in-gap bands tend to be merged, 
Large amounts of the spectral intensity in the upper and lower bands still remain. 
These results imply that the long-range charge alignment is collapsed, 
but a short-range correlation survives. 
%This interpretation was confirmed directly by the charge correlation function.
%This interpretation in $A(q, \omega)$ is confirmed by the charge correlation function $N(q)$; at time $\tau=3/t$, a sharp peak at $q=\pi$ is smeared out and its momentum dependence becomes inconspicuous. 
%
It is worth to note that the band width of the in-gap band is broaden with time evolution [see Figs.~\ref{fig:fig1}(b) and \ref{fig:fig1}(c)]. 
%This will be analyzed in more detail later. 
%
The optical absorption spectra 
$\alpha(\omega)=-(\pi N_L)^{-1} {\rm Im} \langle j (\omega-{\cal H}_{DE}+E+i \gamma)^{-1} j \rangle$ 
are presented in Fig.~\ref{fig:fig1}(d). 
After photon pumping, spectral weight appears inside of the optical gap, and grows up with increasing in time. 
%This spectrum is mainly attributed to the excitation where 
%the relative momentum for the NN electron pair (doublon) and the NN hole one (holon) is changed by the probe photon. 
We confirm, through the analyses in a small cluster system, that 
the lowest component of the in-gap spectral weight 
corresponds to the Drude component in the thermodynamic limit. 
These time evolutions in $A(q, \omega)$ and $\alpha(\omega)$  
are not observed in both the spin-less $V$-$t$ and Hubbard models.  
That is, the localized spins and its coupling with conduction electrons play a central role on these phenomena. 

\begin{figure}[tb]
\begin{center}
\includegraphics[width=0.8\columnwidth,clip]{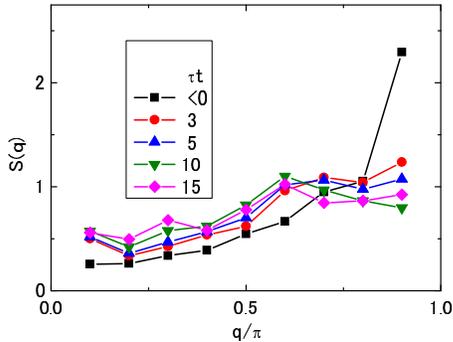}
\caption{(color online) 
Spin-correlation functions at several times.  
}
\label{fig:fig2}
\end{center}
\vspace{-0.6cm}
\end{figure}
Change in the spin sector is monitored by the static spin correlation function, 
$S({\bf q})= \langle {\bf S}_{\bf -q} \cdot {\bf S}_{\bf q} \rangle$,   
where ${\bf S}_q$ is FT of ${\bf S}_i$ [see Fig.~\ref{fig:fig2}]. 
Before pumping,
the spin structure is ferrimagnetic one where the localized spins are aligned antiferromagnetically and the 
total spin is $S^{tot}=\frac{1}{2}+\frac{L+1}{4}$. 
This is seen in large intensity of $S(q \sim \pi)$. 
After pumping, the large AF correlation is rapidly suppressed, 
the momentum dependence in $S(q)$ is almost smeared out, 
%The latter may correspond to the so-called island-type spin structure discussed in the one-dimensional DE model with carrier doping~\cite{garcia}. 
%These results imply that 
and the system is nearly paramagnetic.
% after pumping. 

Time evolutions of both the conduction electron and localized spins 
are summarized in Fig.~\ref{fig:fig3}. 
We plot the second moment $W$ of $A(q, \omega)$ reflecting 
the band width of the in-gap state, the integrated spectral weight $D$ inside of the optical gap in $\alpha(\omega)$, 
and the NN spin correlation $K_S$ for the localized spins. 
These are defined by 
$W=\int^{\omega_U}_{\omega_L} \sum_q A(q, \omega) (\omega-\omega_c)^2 d \omega$ 
with the center of the in-gap band $\omega_c$ and the upper (lower) band edge $\omega_U (\omega_L)$, 
$D=\int^{\omega_U'}_{\omega_L'} \alpha(\omega) d \omega$ with the upper (lower) edge of 
the in-gap component $\omega_U' (\omega_L')$, 
and $K_S=N_B^{-1}\sum_{\langle ij \rangle} \langle {\bf S}_i \cdot {\bf S}_j\rangle$ 
with the number of NN bonds $N_B$.  
An almost identical time dependence is shown in the three curves; they increase linearly after pumping 
and are saturated around $\tau=10/t$. 
The characteristic time scale where $K_S$ shows a shoulder is denoted as $\tau_S$. 
In the inset of Fig.~\ref{fig:fig3}, $K_S$'s for different values of the electron transfer are presented. 
With increasing $\alpha$, a slope of the curve in the region of $\tau < \tau_S$ increases, and the AFM correlation collapses rapidly. 
When $K_S$'s are plotted as functions of $\tau \alpha t$, 
the slopes for several values of $\alpha$ are almost identical.  
We conclude that photo dynamics in the charge and short-range spin sectors are strongly correlated with each other, and are mainly governed by the electron transfer.  
These results are interpreted as follows; motion of the conduction electrons, excited directly by photon,  
destroys the charge and AFM orders. 
%Excess energy flows from the conduction electrons to the localized spin system. 
The collapse of the AFM spin alignment further promotes motion of the conduction electron, and 
finally the two systems steady down cooperatively. 

\begin{figure}[tb]
\begin{center}
\includegraphics[width=0.85\columnwidth,clip]{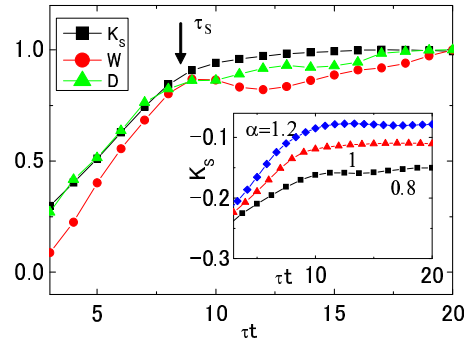}
\caption{(color online) 
Width of the in-gap band in one-particle excitation spectra, $W$, spectral weight inside of the optical gap, $D$, 
and NN spin correlation function, $K_S$. 
Data are subtracted by values at $\tau=0$, and are normalized by 
differences between the minimum and maximum values.   
%Adopted parameter values are the same with those in Fig.~\ref{fig:fig1}. 
We take $(\omega_U, \omega_L)=(6.5, -1)$ and $(\omega_U', \omega_L')=(4, 0.5)$. 
The inset shows $K_S$ for several transfer integrals. 
}
\label{fig:fig3}
\end{center}
\vspace{-0.6cm}
\end{figure}

In the calculations by the ED method explained so far, the spin-angular momentum is conserved. 
%, and  simulations are limited within a time scale of the order of $20/t-30/t$. 
%Before pumping, 
%the ferrimagnetic spin state associated with the antiferromagnetic localized spin alignment is the stable ground state. 
%It is expected, however, in the double exchange model that 
%the lowest photo-excited state in the energy surface, which will be realized in %the long time scale, 
%favors a more spin polarized state, such as a fully polarized ferromagnetic one. 
However, it is expected that, in the simulation for longer time scale, 
the spin relaxation leads the system to the lower excited state on the energy surface.  
Microscopic mechanisms for breaking of the spin conservation in manganites are not clear yet. 
In photo-excited states where the Mn $3d e_g$ orbitals are mainly concerned, 
the orbital angular momentum, ${\bf L}$, is basically quenched. 
%even when the $e_g$ orbital degeneracy is taken into account. 
This is in contrast to the magnetic semiconductors, such as (Ga,Mn)As 
where a large spin-orbit coupling provides a spin relaxation~\cite{chovan}. 
Possible relaxation mechanisms are mixing of the on-site $e_g$ and $t_{2g}$ orbitals by the off-diagonal components of ${\bf L}$, and that induced by lattice distortions with the $\rm T_{\rm 1u}$ symmetry. 
A mixing of the inter-site $e_g$ and $t_{2g}$ orbitals by the GdFeO$_3$-type distortion is 
another candidate. We predict that the photo-induced magnetization is promoted by excitation of the phonons concerning these lattice distortions. 

%Here, we do not touch, anymore, a detailed microscopic mechanism of the spin relaxation. 
We introduce the spin relaxation phenomenologically in the HF formulation,
and examine roles of the spin conservation on time evolution.  
%In the Bloch-type equation given 
In Eq.~(\ref{eq:bloch}), 
we introduce the parallel and perpendicular spin components to the total magnetization ${\bf S} = {\bf e} S$, with a unit vector ${\bf e} $, as~\cite{chovan}
${\bf S}_i={\bf S}_{i \perp}+ S_{i \parallel} {\bf e}$. 
The relaxation terms are given by 
$\partial_\tau S_{i \parallel}|_{relax}=-\Gamma_{L} (S_{i \parallel}-S) $
and 
$\partial_\tau {\bf S}_{i \perp}|_{relax}=-\Gamma_{T} {\bf S}_{i \perp}$, 
where $\Gamma_{L}$ and $\Gamma_{T}$ are the relaxation constants with a relation $\Gamma_{L}=2\Gamma_{T}$. 
A value of $\Gamma_{L(T)}$ should be much small than that for (Ga,Mn)As where 
the spin-relaxation is expected to occur within 100fs. 
We also examine the Landau-Lifshitz-Gilbert-type relaxation $\partial_\tau {\bf S}_{i} |_{relax} = \Gamma_{LG} {\bf S}_i \times \partial_\tau {\bf S}_i$, and confirm that results are not sensitive to the types of the relaxation. 

\begin{figure}[tb]
\begin{center}
\includegraphics[width=0.9\columnwidth,clip]{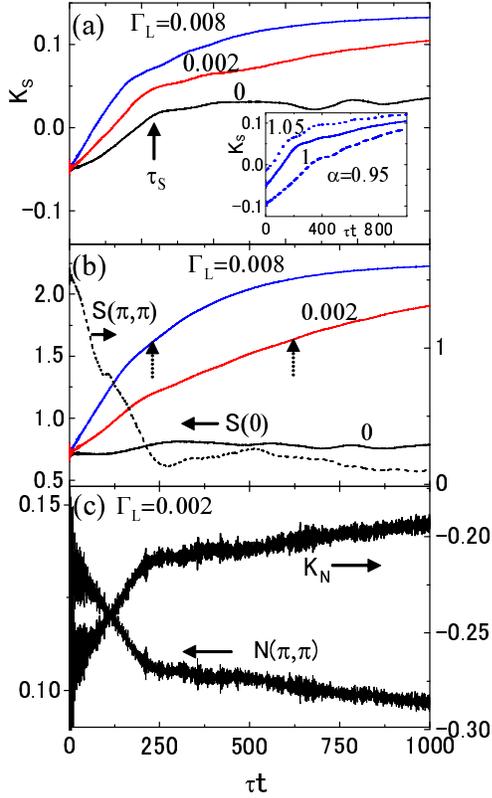}
\caption{(color online) 
(a) NN spin correlation function $K_S$ for several values of 
the relaxation constant $\Gamma_{L}$. 
The inset shows $K_S$ for the several transfer integrals $\alpha t$ and $\Gamma_L=0.002$. 
(b) Spin correlation functions $S(0)$ for several $\Gamma_L$ (bold lines), and 
$S(\pi, \pi)$ for $\Gamma_L=0.002$ (dotted line). 
The dotted arrows represent $\tau_L$.   
(c) Charge correlation function $N(\pi, \pi)$ and NN charge correlation $K_N$. }
\label{fig:fig4}
\end{center}
\vspace{-0.6cm}
\end{figure}
The results by the HF method are presented in Fig.~\ref{fig:fig4}. 
The parameter values are   
$U=8$, $J_H=6$, $V=1$ and $J_S=0.03$. 
We take $\alpha=1$ except for the inset of Fig.~\ref{fig:fig4}(a). 
%Here we take more realistic parameter values for manganites than those in the ED calculation. 
Before pumping, CO associated with the canted-AFM alignment in the localized spins is realized. 
%This difference influences numerical values for the characteristic time scale which will be shown below. 
Two time scales are recognized in the NN spin correlations for the localized spins $K_S$ shown in Fig.~\ref{fig:fig4}(a); a sharp increase in a short-time scale ($\sim 200/t$) depending on the transfer integral [see the inset of Fig.~\ref{fig:fig4}(a)], and a slow increase in a long-time scale depending on $\Gamma_L$. 
The short-time scale corresponds to $\tau_S$ introduced previously. 
The long-time scale termed $\tau_L$ is defined as a time when an amount of change in the FM spin correlation $S(0)$  becomes $1/e$ of its total value; $\tau_L=645/t$ and $250/t$ for $\Gamma_L=0.002$ and $0.008$, respectively. 
% [see Fig.~\ref{fig:fig4}(b)].  
In the case without relaxation, 
a $K_S$ v.s. $\tau$ curve is qualitatively similar with that in the ED method [see Fig.~\ref{fig:fig3}]. 
This implies that with regard to the short-time dynamics with $\Gamma_L=0$,  
a qualitative feature commonly observed by the two methods is reliable.  
A quantitative difference between the $\tau_S$ values in the ED and HF calculations is mainly attributed to the different photon densities $x_{ph}$; 
the present $x_{ph}=L^{-2}=0.01$ is smaller than $x_{ph} \sim 0.1$ in the ED calculation. 
For small $x_{ph}$, a system takes long time to be settled down to a stationary state. 
We checked the above points by the HF calculations in one-dimension; 
a linear increase and a saturation are observed in $K_S$, 
%as the results in Fig.~\ref{fig:fig3} and Fig.~\ref{fig:fig4}(a), 
and $\tau_S$ increases by increasing $L$ from 10 to 30. 

We focus on the results for $\Gamma_L=0.002$. 
As shown in Fig.~\ref{fig:fig4}(b), around $\tau_S$, $S(\pi, \pi)$ disappears, but only a subtle change is seen in $S(0)$; the system becomes almost paramagnetic. 
% around $\tau_S$.  
In the long-time scale beyond $\tau_S$, a remarkable change is seen in $S(0)$ which grows up monotonically toward the full-spin polarization, in contrast to $K_S$ and $S(\pi, \pi)$. 
Charge dynamics, i.e. the long-range charge correlation   
$N({\bf q})= \langle \Delta n_{\bf q} \Delta n_{\bf -q} \rangle$ 
%$N({\bf q})= \langle n_{\bf -q} n_{\bf q} \rangle$, 
and the NN correlation 
$K_N=(2/N_B)\sum_{\langle ij \rangle} \langle \Delta n_i \Delta n_j \rangle$, are shown  
in Fig.~\ref{fig:fig4} (c). We define $\Delta n_i=n_i-1/2$ and its FT, $\Delta n_{\bf q}$. 
%Here we focus mainly on the data for $\Gamma_L=0.002$. 
%Around $\tau_S^{HF}$, $S(\pi, \pi)$ disappears, but only a subtle change is seen in $S(0)$. 
%In brief, the system becomes almost paramagnetic around $\tau_S^{HF}$.  
%In the long-time scale beyond $\tau_S^{HF}$, $S(0)$ grows up monotonically toward the full-spin 
%polarization, but weak changes are seen in $K_S$ and $S(\pi, \pi)$.  
Time dependences of the long- and short-range charge correlations are similar to that in $K_S$, not in $S(0)$. 
These are nearly saturated around $\tau_S$ and do not show a remarkable change beyond $\tau_S$. 
Changes in 
%the long-range FM spin correlation 
$S(0)$ by spin relaxation 
are almost irrelevant to the charge sector. 
This weak correlation between the charge correlations and $S(0)$, corresponding to magnetization, 
is not expected from 
%a conventional picture in 
the equilibrium state where these are combined strongly. 
The key ingredient in this dynamical separation between the spin and charge sectors is spin conservation; magnetization dynamics is governed by the spin relaxation rather than the charge motion. 
The present case is in contrast to the photo-induced dynamics in electron-phonon systems  
where conserved quantities in a phonon sector does not exist. 

%In conclusion, 
%we present a theory of photo-induced dynamics in the DE model 
%by utilizing the two complimentary calculation methods. 
%The two-time scales, governed by the electron transfer and the spin relaxation, characterize the photo-induced phenomena. 
%that spin and charge dynamics are characterized by the two-time scales. 
%In the short-time scale, governed by the electron-transfer integral, 
%the spin and charge dynamics evolve cooperatively. 
%On the contrary, in the long-time one characterized by the spin relaxation, 
%the charge sector does not follow a remarkable change in the FM correlation.  
%
From the calculated results, we provide a whole picture of the photo-induced spin and charge dynamics. 
In the short-time scale, charge motions governed by the electron transfer 
destroy the CO and short-range AFM correlations cooperatively. 
Excess energy given by the pump photon flows from the 
conduction electrons to the localized spins. 
This state is further relaxed through the spin relaxation in long-space and long-time scales.
Changes in the charge sector is almost saturated in the short-time scale 
and do not follow the spin relaxation. 
This scenario provides a microscopic interpretation for the experimentally observed 
multiple time scales in $\Delta R$ and $\Delta \theta$. 
The present dynamical coupling/separation in the multiple degrees of freedom 
is not expected from a conventional picture in equilibrium state and are key issues to reveal the photo-induced dynamics in correlated system. 

\par
Authors would like to thank K.~Satoh, 
%K.~Miyano, H.~Okamoto, Y.~Tokura, S.~Iwai, A.~Takahashi, K.~Yonemitsu, 
K.~Nasu and T.~Arima for their valuable discussions. 
This work was supported by JSPS KAKENHI, TOKUTEI from MEXT, and Grand challenges in next-generation integrated nanoscience. 

\vfill \eject

\vfill
\eject

\begin{references}
\bibitem{photon}
A Lorentzian factor in Eq.~(\ref{eq:psitau}) comes from an expansion coefficient in the time-dependent 
perturbation theory, given by 
$c_n(\tau)=\langle n | j | 0\rangle \int^{\tau}_{-\infty} d \tau' A_{pump}(\tau') e^{i(E_n-E_0) \tau'}$ 
with the $n$-th eigen state $ | n \rangle$ and the eigen energy $E_n$. 
By substituting a damped oscillator form of $A_{pump}(\tau')$ and integrating out under a condition of $ \tau >> \gamma_0^{-1}$, we obtain $|\Psi(\tau) \rangle$ in Eq.~(\ref{eq:psitau}).
\bibitem{approxi1}
After pumping, we introduce the approximation that the eigen energies in the denominator are replaced by $\langle \Psi(\tau) |{\cal H}_{DE}| \Psi (\tau) \rangle$. 
Average difference between the optical absorption data with and without this approximation 
is less than 10 percent.
\end{references}
\end{document}